\documentclass{article}
\usepackage{amssymb}
\usepackage{txfonts}
\title{\bf Construction of Hypergeometric Solutions to the $q$-Painlev\'e Equations}
\author{K.~Kajiwara${}^{1,3}$, T.~Masuda${}^2$, M.~Noumi${}^2$, 
Y.~Ohta${}^2$ and Y.~Yamada${}^2$\\[3mm]
{\normalsize ${}^1$ Graduate School of Mathematics, Kyushu University,
6-10-1 Hakozaki, Fukuoka 812-8581, Japan}\\
{\normalsize ${}^2$ Department of Mathematics, Kobe University, Rokko,
Kobe 657-8501, Japan}\\
{\normalsize ${}^3$ School of Mathematics and Statistics F07, University
of Sydney, Sydney, NSW 2006, Australia}
}
\date{January 30, 2005}
\newtheorem{thm}{Theorem}[section]
\newtheorem{prop}[thm]{Proposition}

\newcommand{\ds}[1]{{\displaystyle #1}}

\newcommand{\dfrac}[2]{{\displaystyle\frac{#1}{#2}}}
\newcommand{\ol}{\overline}
\newcommand{\ul}{\underline}

\newcommand{\og}{\overline {g}}
\newcommand{\ot}{\overline {t}}

\def\df{\underline {f}}

\def\dt{\underline {t}}
\def\lra{\leftrightarrow}
\def\hf{\frac{1}{2}}

\begin{document}
\maketitle
\begin{abstract}
Hypergeometric solutions to the $q$-Painlev\'e equations
are constructed by direct linearization of disrcrete
Riccati equations.  The decoupling factors are explicitly
determined so that the linear systems give rise to
$q$-hypergeometric equations.
\end{abstract}

\section{Introduction}
This article is a continuation of our previous work \cite{KMNOY:HGqPL}
on the hypergeometric solutions to the $q$-Painlev\'e equations in the
following degeneration diagram of affine Weyl group symmetries
\cite{Sakai,RGTT} :
\begin{equation}
 \begin{array}{cccccccccccccc}
E_{8}^{(1)} & \rightarrow & E_{7}^{(1)} & \rightarrow &
E_{6}^{(1)} & \rightarrow & D_{5}^{(1)} & \rightarrow &
A_{4}^{(1)} & \rightarrow & (A_2+A_1)^{(1)} & \rightarrow &
(A_1+{{\displaystyle A_1}\atop \hbox{${\scriptstyle |\alpha|^2=14}$}})^{(1)}
\end{array}\label{fig:q-Pcoales}
\end{equation}
The list of $q$-Painlev\'e equations we are going to investigate is
given in section 3 below.  We remark that these $q$-Painlev\'e equations
were discovered through various approaches to discrete Painlev\'e
equations, including singularity confinement analysis, compatibility
conditions of linear difference equations, affine Weyl group symmetries
and $\tau$-functions on the lattices.  Also, in Sakai's framework
\cite{Sakai}, each of these $q$-Painlev\'e equations is constructed in a
unified manner as the birational action of a translation of the
corresponding affine Weyl group on a certain family of rational
surfaces.

\par\medskip In \cite{KMNOY:elliptic} we have introduced the formulation
of discrete Painlev\'e equations based on the geometry of plane
curves on $\mathbb{P}^2$.  On that basis we were able in the first part
of \cite{KMNOY:HGqPL} to find suitable coordinates for linearization of
the $q$-Painlev\'e equations into three-term relations of hypergeometric
functions.  As a result we obtained the following degeneration diagram
of basic hypergeometric functions corresponding to
(\ref{fig:q-Pcoales}):
\begin{displaymath}
 \begin{array}{cccccccccccccc}
{\displaystyle {{\rm balanced}\atop{{}_{10}W_9}}} & \rightarrow
& {\displaystyle {{}_8W_7}}& \rightarrow
& {\displaystyle {{\rm balanced}\atop{{}_{3}\varphi_2}}} & \rightarrow
& {\displaystyle {}_2\varphi_1}& \rightarrow
& {\displaystyle {}_1\varphi_1} & \rightarrow
& {\displaystyle {{}_1\varphi_1\left(\begin{array}{c}a\\0 \end{array};q,z\right)}}\atop
{\displaystyle {{}_1\varphi_1\left(\begin{array}{c}0\\b \end{array};q,z\right)}} & \rightarrow
& {\displaystyle {}_1\varphi_1\left(\begin{array}{c}0\\-q \end{array};q,z\right)}
\end{array}
\end{displaymath}
This shows the usefulness of geometric consideration in the study of
particular solutions of discrete Painle\'e equations.
In order to determine explicit solutions to those equations written in
the literature, further steps of precise computations are required since
the variables to be solved are fixed in advance.  We have shown only the
results for them in the second part of \cite{KMNOY:HGqPL}.  The purpose
of this article is to present these explicit solutions including subtle
gauge factors and to show the calculation in detail based on the direct
linearization of discrete Riccati equations.

\par\medskip
This article is organized as follows: In section 2, we explain the
procedure to construct hypergeometric solutions through the
linearization of discrete Riccati equations. We demonstrate this
procedure by taking the case of $E_7^{(1)}$ as an example. In section 3,
we give the list of $q$-Painlev\'e equations and their hypergeometric
solutions, with the data that are necessary for constructing
solutions. Among the cases in the diagram (\ref{fig:q-Pcoales}) we have excluded the case
of $D_5^{(1)}$, since a complete identification of hypergeometric solutions
has already been given in \cite{Sakai:qp6sol}.
\section{Construction of Hypergeometric Solutions}
\subsection{Discrete Riccati Equation and Its Linearization}
The $q$-Painlev\'e equations admit particular solutions characterized by
discrete Riccati equations for special values of parameters. Reduction
to discrete Riccati equation has been already done for all the
$q$-Painlev\'e equations\cite{RGTT,MSY,KNY:qp4,KK:qp3}. We also note
that such special situations have clear geometrical meaning, as was
discussed in \cite{KMNOY:HGqPL}.  The basic idea for constructing
hypergeometric solution is as follows: we linearize the discrete Riccati
equations to yield second order linear $q$-difference equations. We then
identify them with the three-term relation of an appropriate basic
hypergeometric series.

Let us explain this procedure in detail. Suppose we have
a discrete Riccati equation of the form 
\begin{equation}
 \overline{z}=\frac{Az+B}{Cz+D},\label{riccati:general}
\end{equation}
where $z=z(t)$ and $\overline{z}=z(qt)$. We also use the notation
$\underline{z}=z(t/q)$, and so forth. Moreover, the coefficients $A$,
$B$, $C$ and $D$ are functions of $t$. First let us put an ansatz 
\begin{equation}\label{ansatz1:general}
 z=\frac{F}{G}.
\end{equation}
Then the discrete Riccati equation is linearized to
\begin{equation}\label{linear:general}
\dfrac{\overline{F}}{H}=AF+BG,\quad \dfrac{\overline{G}}{H}=CF+DG,
\end{equation}
where $H$ is an arbitrary decoupling factor. Eliminating $G$ from
eq.(\ref{linear:general}) we have for $F$ the three-term relation 
\begin{equation}\label{3termF:general}
 \overline{F}+c_1F+c_2\underline{F}=0,\quad c_1=-\frac{H}{\underline{B}}(A\underline{B}+B\underline{D}),
\quad c_2=\frac{B}{\underline{B}}H\underline{H}(\underline{A}\underline{D}-\underline{B}\underline{C}).
\end{equation}
The three-term relation for a basic hypergeometric series often takes
the form 
\begin{equation}\label{3termPhi:general}
 V_1(\overline{\Phi}-\Phi)+V_2\Phi+V_3(\underline{\Phi}-\Phi)=0, 
\end{equation}
where the coefficients $V_1,V_2$ and $V_3$ are factorized into binomials
involving the independent variable and parameters. Comparing
eqs.(\ref{3termF:general}) with (\ref{3termPhi:general}), we have
\begin{equation}
 \frac{V_2}{V_1}=1+c_1+c_2,\quad \frac{V_3}{V_2}=c_2.
\end{equation}
We look for the decoupling factor $H$ so that these quantities
factorize. We then identify the three-term relation with that for
appropriate hypergeometric function.  This is done by trial and error
with the aid of computer algebra, but it is not practically difficult
since we already know the hypergeometric function and its three-term
relation to appear for each $q$-Painlev\'e equation.  

\par\medskip

\noindent\textbf{Step 1.}  
\textit{Find the decoupling factor $H$ such that
\begin{equation}
  \frac{V_2}{V_1}=1+c_1+c_2,\quad \frac{V_3}{V_2}=c_2,
\end{equation}
factorize. Then identify the three-term relation
\begin{equation}
  V_1(\overline{F}-F)+V_2F+V_3(\underline{F}-F)=0,
\end{equation}
with that for an appropriate hypergeometric function.}

\par\medskip

Similarly, we have for $G$ the three-term relation 
\begin{equation}\label{3termG:general}
 \overline{G}+\tilde d_1G+\tilde d_2\underline{G}=0,
\quad \tilde d_1=-\frac{H}{\underline{C}}(D\underline{C}+C\underline{A}),
\quad \tilde d_2=\frac{C}{\underline{C}}H\underline{H}(\underline{A}\underline{D}
-\underline{B}\underline{C}).
\end{equation}
However, usually $1+\tilde d_1+\tilde d_2$ does not factorize for $H$
obtained above. Replacing $G$ with $\kappa
G~(\overline{\kappa}=k\kappa)$, we have 
\begin{equation}
 z=\frac{1}{\kappa}\frac{F}{G},
\end{equation}
\begin{equation}\label{3termG2:general}
 \overline{G}+d_1G+d_2\underline{G}=0,\quad d_1=-\frac{H}{k\underline{C}}(D\underline{C}+C\underline{A}),
\quad d_2=\frac{C}{\underline{C}}\frac{H\underline{H}}{k\underline{k}}
(\underline{A}\underline{D}-\underline{B}\underline{C}).
\end{equation}
We then look for $k$ so that $1+d_1+d_2$ factorizes, and identify
eq.(\ref{3termG2:general}) with the three-term relation of the same
hypergeometric function as $F$ with different parameters. Putting
$H/k=K$, this is equivalent to the follwing procedure:

\par\medskip

\noindent\textbf{Step 2.} 
\textit{In the three-term relation 
\begin{equation}
 \overline{G}+d_1G+d_2\underline{G}=0,\quad d_1=-\frac{K}{\underline{C}}(D\underline{C}+C\underline{A}),
\quad d_2=\frac{C}{\underline{C}}K\underline{K}(\underline{A}\underline{D}-\underline{B}\underline{C}),
\end{equation}
find decoupling factor $K$ so that 
\begin{equation}
  \frac{U_2}{U_1}=1+d_1+d_2,\quad \frac{U_3}{U_2}=d_2,
\end{equation}
factorize. Then identify the three-term relation with that for an
appropriate hypergeometric function. Now we have 
\begin{equation}
 z=\frac{1}{\kappa}~\frac{\theta_1\Phi}{\theta_2\Psi},\quad \frac{H}{K}=k,\quad 
\frac{\overline{\kappa}}{\kappa}=k,
\end{equation}
where $\Phi$ and $\Psi$ are some hypergeometric functions, and
$\theta_i$ ($i=1,2$) are constants(gauge factors).}
\par\medskip
Finally we determine the gauge factors $\theta_1$ and $\theta_2$:
\par\medskip
\noindent\textbf{Step 3.} \textit{Compare the linear relations,
\begin{equation}
\frac{\theta_1\overline{\Phi}}{H}=A\theta_1\Phi+\kappa
 B\theta_2\Psi,\quad \frac{\overline{\kappa}\theta_2\overline{\Psi}}{H}=C\theta_1\Phi+\kappa D\theta_2\Psi,
\label{riccati:linear}
\end{equation}
with contiguity relations of the hypergeometric functions to determine
$\theta_1$ and $\theta_2$.}

\subsection{An Example: Case of $E_7^{(1)}$}
In this section we demonstrate the construction of the hypergeometric
solution to the $q$-Painlev\'e equation of type $E_7^{(1)}$ as an
example, following the procedure in the previous section.

Before proceeding, let us first summarize the definition and terminology
of the basic hypergeometric series~\cite{Gasper-Rahman}. The basic
hypergeometric series ${}_{r}\varphi_{s}$ is given by 
\begin{equation}
\begin{array}{c}
{}_{r}\varphi_s
\left(
\begin{array}{c}
a_1,\ldots,a_r\\
b_1,\ldots, b_s
\end{array};q,z
\right)=
\ds{\sum_{n=0}^{\infty}}
\dfrac{(a_1;q)_n\cdots (a_r;q)_n}
      {(b_1;q)_n\cdots (b_s;q)_n (q;q)_n}
\left[(-1)^n
      q^{\left(n\atop 2\right)}
\right]^{1+s-r}z^n,\\[12pt]
\qquad(a;q)_n=(1-a)(1-qa)\cdots(1-q^{n-1}a). 
\end{array}
\end{equation}
The basic hypergeometric series ${}_{r+1}\varphi_{r}$ is called
\textit{balanced}\footnote{The ``balanced ${}_{3}\varphi_2$'' in the
diagram (\ref{fig:q-Pcoales}) is due to the convention that was used in
~\cite{Gupta-Ismail-Masson}.}
if the condition
\begin{equation}
qa_1a_2\cdots a_{r+1}=b_{1}b_2\cdots b_{r},\quad z=q,
\end{equation}
is satisfied, and is called \textit{very-well-poised} if the condition 
\begin{equation}
qa_1=a_2b_1=\cdots =a_{r+1}b_r,\quad 
a_2=qa_1^{\hf},\quad a_3=-qa_1^{\hf},\label{very-well-poised}
\end{equation}
is satisfied. A very-well-poised hypergeometric series
${}_{r+1}\varphi_r$ is denoted as ${}_{r+1}W_r$:
\begin{equation}
{}_{r+1}W_r(a_1;a_4,\ldots,a_{r+1};q,z)=
{}_{r}\varphi_s\left(
\begin{array}{c}
a_1,qa_1^{\hf},-qa_1^{\hf},a_4\ldots,a_{r+1}\\
a_1^{\hf},-a_1^{\hf},qa_1/a_4,\ldots,qa_1/a_{r+1} 
\end{array};q,z\right).
\end{equation}

Now the $q$-Painlev\'e equation of type $E_7^{(1)}$ is given
by~\cite{Sakai,RGTT}
\begin{equation}
\begin{array}{l}
\medskip
\dfrac{(f\og - \ot t)(fg-t^2)}{(f\og-1)(fg-1)} 
=\dfrac{(f-b_1t)(f-b_2t)(f-b_3t)(f-b_4 t)}
       {(f-b_5)(f-b_6)(f-b_7)(f-b_8)},\\
\dfrac{(fg - t^2)(\df g-t\dt)}{(fg-1)(\df g-1)}
=\dfrac{\left(g-\dfrac{t}{b_1}\right)\left(g-\dfrac{t}{b_2}\right)
        \left(g-\dfrac{t}{b_3}\right)\left(g-\dfrac{t}{b_4}\right)}
       {\left(g-\dfrac{1}{b_5}\right)\left(g-\dfrac{1}{b_6}\right)
        \left(g-\dfrac{1}{b_7}\right)\left(g-\dfrac{1}{b_8}\right)},
\end{array}
\label{eqn:e7}
\end{equation} 
where $t$ is the independent variable and $b_i$ ($i=1,\ldots,8$) are
parameters satisfying
\begin{equation}
\ot=qt,\quad b_1b_2b_3b_4=q,\quad b_5b_6b_7b_8=1.\label{param:cond1}
\end{equation}

\begin{prop}~\cite{RGTT,MSY}
In case of $b_1b_3=b_5b_7$, eq.(\ref{eqn:e7}) admits a specialization to
the discrete Riccati equation,
\begin{eqnarray}
\og &=& \frac{(t\ot -1)f + t\left\{-(b_6+b_8)\ot + (b_2+b_4)\right\}}
{\left\{-(b_6+b_8)+(b_2+b_4)t\right\} f+b_6b_8(1-t\ot)},\label{e7:ric1}\\[3mm]
f &=& \frac{(t^2-1)b_5b_7g+t\left\{(b_1+b_3)-(b_5+b_7)t\right\}}
{\left\{t(b_1+b_3)-(b_5+b_7)\right\}g+(1-t^2)}.\label{e7:ric2}
\end{eqnarray}
\end{prop}

\noindent 
As was pointed out in~\cite{KMNOY:HGqPL}, in the cases of
$E_{6,7,8}^{(1)}$ the variables $f$ and $g$ are not expressed by ratio
of hypergeometric functions. We choose the variable as\footnote{This variable 
$z$ should be understood as a ratio of $\tau$ functions.}
\begin{equation}
z=\frac{g-t/b_1}{g-1/b_5}.
\end{equation}
Then the discrete Riccati equation (\ref{e7:ric1}) and (\ref{e7:ric2})
is rewritten as 
\begin{displaymath}
\overline{z}=\frac{Az+B}{Cz+D},
\end{displaymath}
with 
\begin{eqnarray}
A&=&b_1b_5(-b_3+b_5 t)\nonumber\\
&&\times\biggl[
b_4b_6b_8q^2t^3+(b_1b_4b_5 - b_4^2b_5 - b_1b_4b_6-b_1b_4b_8-b_5b_6b_8q)qt^2\nonumber\\
&&\quad +(b_1b_4^2 + b_4b_5b_6q + b_4b_5b_8q+b_1b_6b_8q-b_4b_6b_8q)t -b_1b_4b_5
\biggr],\label{A}\\
B&=&-(b_1-b_4)b_5^2(b_1b_4-qb_6b_8)t(b_5-b_3 t)(-1+qt^2),\label{B}\\
C&=&-b_1^2b_4(b_5-b_6)(b_5-b_8)(b_3-b_5t)(-1+qt^2),\label{C}\\
D&=&b_1b_5(b_5-b_3t)\nonumber\\
&&\times\biggl[
-b_1b_4b_5qt^3 + (b_1b_4^2+b_4b_5b_6q+b_4b_5b_8q+b_1b_6b_8q-b_4b_6b_8q)t^2\nonumber\\
&&\quad + (b_1b_4b_5-b_4^2b_5-b_1b_4b_6-b_1b_4b_8-b_5b_6b_8q)t + b_4b_6b_8
\biggr].\label{D}
\end{eqnarray}

\noindent
\textbf{Step 1.} We choose the decoupling factor $H$ as
\begin{equation}
H=\frac{1}{qb_1 b_5 (b_5 t-b_3)(b_1t-b_5)(b_4 t-b_6)(b_4 t-b_8)}. 
\end{equation}
Then we have for $F$ the three-term relation 
\begin{equation}
\begin{array}{l}
\medskip
V_1(\overline{F}-F)+V_2F+V_3(\underline{F}-F)=0,\\
\medskip
\dfrac{V_2}{V_1}
=\dfrac{b_5(b_3-b_4)(b_1b_4-qb_6b_8)(-1+t)(1+t)(-1+qt^2)}
       {q(b_5-b_1t)(-b_6+b_4 t)(-b_8+b_4t)(-b_3+b_5t)},\\
\dfrac{V_3}{V_1}
=\dfrac{(b_5-b_3 t)(-b_1+b_5 t)(-b_4+b_6 t)(-b_4+b_8t)(-1+qt^2)}
       {(b_5-b_1 t)(-b_6+b_4 t)(-b_8+b_4 t)(-b_3+b_5 t)(-q+t^2)}.
\end{array}\label{F:3-term}
\end{equation}
The three-term relation for the very-well-poised basic hypergeometric
series
\begin{equation}
\Phi = {}_8W_7\left(a_0;a_1,a_2,a_3,a_4,a_5;q,\frac{q^2a_0^2}{a_1a_2a_3a_4a_5}\right),
\label{8W7:definition}
\end{equation}
is given by~\cite{Ismail-Rahman}
\begin{equation}
\begin{array}{l}
U_1(\overline{\Phi}-\Phi)+U_2\Phi + U_3(\underline{\Phi}-\Phi)=0,\qquad 
\overline{\Phi}=\Phi\,|_{a_2\rightarrow qa_2,a_3\rightarrow a_3/q}, \quad 
\underline{\Phi}=\Phi\,|_{a_2\rightarrow a_2/q,a_3\rightarrow qa_3},\\ [4mm]
U_1
=\dfrac{(1-a_2)\left(1-\dfrac{a_0}{a_2}\right)\left(1-\dfrac{qa_0}{a_2}\right)
        \left(1-\dfrac{qa_0}{a_1a_3}\right)
        \left(1-\dfrac{qa_0}{a_3a_4}\right)
        \left(1-\dfrac{qa_0}{a_3a_5}\right)}
       {a_3\left(1-\dfrac{a_2}{a_3}\right)\left(1-\dfrac{qa_2}{a_3}\right)},
\quad U_3=U_1|\,_{a_2\lra a_3}, \\[4mm]
U_2=\dfrac{qa_0^2}{a_1a_2a_3a_4a_5}
    \left(1-\dfrac{qa_0}{a_2a_3}\right)(1-a_1)(1-a_4)(1-a_5). 
\end{array}\label{8W7:3-term}
\end{equation} 
Comparing eqs.(\ref{F:3-term}) with (\ref{8W7:3-term}), we identify $F$
with ${}_8W_7$ as 
\begin{equation}
F\propto 
{}_8W_7\left(\frac{b_1b_8}{b_3b_5};
\frac{qb_8}{b_5},\frac{b_1t}{b_5},\frac{b_1}{b_5t},\frac{b_2}{b_3},\frac{b_4}{b_3};
q,\dfrac{b_5}{b_6}\right).
\end{equation}

\noindent
\textbf{Step 2.} We choose the decoupling factor $K$ as
\begin{equation}
K=\frac{1}{b_1 b_5 (qb_5 t-b_3)(b_1 t-b_5)(b_4 t-b_6)(b_4 t-b_8)}.
\end{equation}
Then we have for $G$ the three-term relation 
\begin{equation}
\begin{array}{l}
\medskip
X_1(\overline{G}-G)+X_2F+X_3(\underline{G}-G)=0,\\
\medskip
\dfrac{X_2}{X_1}
=\dfrac{b_5(b_3-b_4)(b_1b_4-b_6b_8)(-1+t)(1+t)(-1+qt^2)}
       {(b_5-b_1t)(-b_6+b_4 t)(-b_8+b_4t)(-b_3+qb_5t)},\\
\dfrac{X_3}{X_1}
=\dfrac{(qb_5-b_3 t)(-b_1+b_5 t)(-b_4+b_6 t)(-b_4+b_8t)(-1+qt^2)}
       {(b_5-b_1 t)(-b_6+b_4 t)(-b_8+b_4 t)(-b_3+qb_5 t)(-q+t^2)}.
\end{array}\label{G:3-term}
\end{equation}
Comparing eqs.(\ref{G:3-term}) with (\ref{8W7:3-term}), we have
\begin{equation}
G\propto {}_8W_7 
\left(\frac{b_1b_8}{b_3b_5}
      ;\frac{b_8}{b_5},\frac{b_1t}{b_5},\frac{b_1}{b_5t},\frac{b_2}{b_3},\frac{b_4}{b_3}
      ;q,\frac{qb_5}{b_6}\right). 
\end{equation}
Moreover, from $k=H/K=(1-b_3/qb_5t)/(1-b_3/b_5t)$, we have
$\kappa=1-b_3/b_5t$. Therefore we obtain 
\begin{equation}
z\propto\frac{1}{1-\dfrac{b_3}{b_5t}}~
\dfrac{{}_8W_7\left(a_0;qa_1,a_2,a_3,a_4,a_5;q,\dfrac{q  a_0^2}{a_1a_2a_3a_4a_5}\right)}
      {{}_8W_7\left(a_0, a_1,a_2,a_3,a_4,a_5;q,\dfrac{q^2a_0^2}{a_1a_2a_3a_4a_5}\right)},
\end{equation}
with 
\begin{equation}
a_0=\frac{b_1b_8}{b_3b_5},\quad 
a_1=\frac{b_8}{b_5},\quad a_2=\frac{b_1t}{b_5}, \quad 
a_3=\frac{b_1}{b_5t},\quad a_4=\frac{b_2}{b_3},\quad a_5=\frac{b_4}{b_3},\quad 
\frac{q^2a_0^2}{a_1a_2a_3a_4a_5}=\dfrac{qb_5}{b_6}. 
\label{e7:param}
\end{equation}

\noindent
\textbf{Step 3.} Let us put
\begin{equation}
F=\theta(qa_1,a_2,a_3)\Phi(qa_1,a_2,a_3),\quad 
G=\theta(a_1,a_2,a_3)\Phi(a_1,a_2,a_3),  \label{FG}
\end{equation}
where $\theta(a_1,a_2,a_3)$ is a gauge factor to be determined. 
Here, we have omitted the dependence of $a_0,a_4$ and $a_5$, since they
are not relevant to the calculation. Then linear equations
(\ref{riccati:linear}) yield
\begin{equation}
\frac{1}{\kappa HB}
\frac{\theta(qa_1,qa_2,a_3/q)}{\theta(a_1,a_2,a_3)}~\Phi(qa_1,qa_2,a_3/q)
=\frac{A}{\kappa B}
\frac{\theta(qa_1,a_2,a_3)}{\theta(a_1,a_2,a_3)}~\Phi(qa_1,a_2,a_3)+\Phi(a_1,a_2,a_3),
\label{e7:linear1}
\end{equation}
and 
\begin{equation}
\frac{\overline{\kappa}}{\kappa HD}
\frac{\theta(a_1,qa_2,a_3/q)}{\theta(a_1,a_2,a_3)}~\Phi(a_1,qa_2,a_3/q)
=\frac{C}{\kappa D}
\frac{\theta(qa_1,a_2,a_3)}{\theta(a_1,a_2,a_3)}~\Phi(qa_1,a_2,a_3)+\Phi(a_1,a_2,a_3),
\label{e7:linear2}
\end{equation}
respectively. Now, we have the contiguity relations for
$\Phi={}_8W_7$~\cite{Ismail-Rahman}
\begin{equation}
\begin{array}{l}
\dfrac{a_1\left(1-\dfrac{a_0q}{a_1a_3}\right)\left(1-\dfrac{a_0q}{a_1a_4}\right)
          \left(1-\dfrac{a_0q}{a_1a_5}\right)}
      {1-\dfrac{a_0q}{a_1}}~\Phi(a_1/q,a_2,a_3)
-\Big(a_1 \lra a_2\Big)\\[6mm]
=(a_1-a_2)\left(1-\dfrac{a_0^2q^2}{a_1a_2a_3a_4a_5}\right)\Phi(a_1,a_2,a_3),
\end{array}   \label{e7:CR1}
\end{equation}
\begin{equation}
\begin{array}{l}
(a_2-1)\left(1-\dfrac{a_0}{a_2}\right)\Phi(a_1/q,qa_2,a_3)
+\left(1-\dfrac{a_1}{q}\right)\left(1-\dfrac{a_0q}{a_1}\right)
\Phi(a_1,a_2,a_3)\\[3mm]
=\left(a_2-\dfrac{a_1}{q}\right)\left(1-\dfrac{a_0q}{a_1a_2}\right)
\Phi(a_1/q,a_2,a_3).
\end{array}   \label{e7:CR2}
\end{equation}
We denote eqs.(\ref{e7:CR1}) and (\ref{e7:CR2}) as CR1$[a_1,a_2,a_3]$
and CR2$[a_1,a_2,a_3]$, respectively. Moreover, note that the relations
CR1$[a_1,a_2,a_3]$ and CR2$[a_1,a_2,a_3]$ hold for any permutation of
$a_1,a_2$ and $a_3$, since these parameters are on equal footing in
$\Phi={}_8W_7$. 

Now we eliminate $\Phi(a_1,a_2,a_3/q)$ from CR1$[a_1,a_3,a_2]$ and
CR2$[a_3,a_2,a_1]$. Shifting $a_1$ to $qa_1$, we have a linear relation
among $\Phi(a_1,qa_2,a_3/q)$, $\Phi(a_1,a_2,a_3)$ and
$\Phi(a_1/q,a_2,a_3)$, which should coincide with
eq. (\ref{e7:linear1}). Similarly, eliminating $\Phi(qa_1,a_2,a_3/q)$
from CR1$[qa_1,a_3,a_2]$ and CR2$[a_3,a_1,a_2]$, we have a linear
relation among $\Phi(qa_1,a_2,a_3),\Phi(a_1,a_2,a_3)$ and
$\Phi(a_1,a_2,a_3/q)$. Elimination further $\Phi(a_1,a_2,a_3/q)$ from
this relation and CR2$[a_3,a_2,a_1]$ yields a linear relation among
$\Phi(a_1,qa_2,a_3/q),\Phi(qa_1,a_2,a_3)$ and $\Phi(a_1,a_2,a_3)$, which
should coincide with eq.(\ref{e7:linear2}). From these calculations, we
find that $\theta(a_1,a_2,a_3)$ should satisfy
\begin{equation}
\begin{array}{l}
\smallskip
\dfrac{\theta(a_1,qa_2,a_3/q)}{\theta(a_1,a_2,a_3)}=
\dfrac{\left(1-\dfrac{a_2}{a_0}\right)\left(1-\dfrac{a_3a_5}{a_0q}\right)}
      {\left(1-\dfrac{a_2a_5}{a_0}\right)\left(1-\dfrac{a_3}{a_0q}\right)}=
\dfrac{\left(1-\dfrac{b_3t}{b_8}\right)\left(1-\dfrac{b_4}{b_8qt}\right)}
      {\left(1-\dfrac{b_4t}{b_8}\right)\left(1-\dfrac{b_3}{b_8qt}\right)}, \\
\smallskip
\dfrac{\theta(qa_1,a_2,a_3)}{\theta(a_1,a_2,a_3)}
=1-\dfrac{a_1}{a_0}=1-\dfrac{b_3}{b_1}, \\
\dfrac{\theta(qa_1,qa_2,a_3/q)}{\theta(a_1,a_2,a_3)}=
\dfrac{\theta(a_1,qa_2,a_3/q)}{\theta(a_1,a_2,a_3)}
\times \dfrac{\theta(qa_1,a_2,a_3)}{\theta(a_1,a_2,a_3)},
\end{array}
\end{equation}
which yield
\begin{equation}
\theta(a_1,a_2,a_3)
=\frac{\left(\dfrac{a_2}{a_0q},\dfrac{a_3a_5}{a_0q},\dfrac{a_1}{a_0q}\right)_\infty}
      {\left(\dfrac{a_2a_5}{aq},\dfrac{a_3}{a_0q}\right)_\infty} 
=\frac{\left(\dfrac{b_3t}{qb_8},\dfrac{b_4}{qb_8t},\dfrac{b_3}{qb_1}\right)_\infty}
      {\left(\dfrac{b_4t}{qb_8},\dfrac{b_3}{qb_8t}\right)_\infty}. 
\end{equation}
Therefore we arrive at the final result 
\begin{equation}
\begin{array}{ll}
z&=\dfrac{1}{1-\dfrac{b_3}{b_5t}}~
   \dfrac{\theta(qa_1,a_2,a_3)}{\theta(a_1,a_2,a_3)}~
   \dfrac{\Phi(qa_1,a_2,a_3)}{\Phi(a_1,a_2,a_3)}\\[7mm]
&=\dfrac{1-\dfrac{b_3}{b_1}}{1-\dfrac{b_3}{b_5t}}~
  \dfrac{{}_8W_7\left(a_0;qa_1,a_2,a_3,a_4,a_5;q,
                      \dfrac{q  a_0^2}{a_1a_2a_3a_4a_5}\right)}
        {{}_8W_7\left(a_0, a_1,a_2,a_3,a_4,a_5;q,
                      \dfrac{q^2a_0^2}{a_1a_2a_3a_4a_5}\right)}. 
\end{array}
\end{equation}
\section{Hypergeometric Solutions}
Hypergeometric solutions to other $q$-Painlev\'e equations can be
constructed by the same procedure as that was demonstrated in the
previous section. Instead of describing full procedure, we give a list
of equations, solutions and the other data that are necessary for
construction of solutions. We note that the case of $D_5^{(1)}$ is
omitted as mentioned in the introduction.
\subsection{Case of $E_8^{(1)}$}
\subsubsection{Equation and Solution}
\begin{enumerate}
 \item $q$-Painlev\'e equation \cite{RGTT,MSY,ORG}
\begin{equation}\label{qPe8}
\begin{array}{l}
\dfrac{(\ol{g}\ol{s}t-f)(gst-f)-(\ol{s}^2t^2-1)(s^2t^2-1)}
      {\left(\dfrac{\ol{g}}{\ol{s}t}-f\right)\left(\dfrac{g}{st}-f\right)
      -\left(1-\dfrac{1}{\ol{s}^2t^2}\right)\left(1-\dfrac{1}{s^2t^2}\right)}
=\dfrac{P(f,t,m_1,\ldots,m_7)}{P(f,t^{-1},m_7,\ldots,m_1)},  \\[8mm]
\dfrac{(\ul{f}s\ul{t}-g)(fst-g)-(s^2\ul{t}^2-1)(s^2t^2-1)}
      {\left(\dfrac{\ul{f}}{s\ul{t}}-g\right)\left(\dfrac{f}{st}-g\right)
      -\left(1-\dfrac{1}{s^2\ul{t}^2}\right)\left(1-\dfrac{1}{s^2t^2}\right)}
=\dfrac{P(g,s,m_7,\ldots,m_1)}{P(g,s^{-1},m_1,\ldots,m_7)}, 
\end{array}
\end{equation}
where 
\begin{equation}
\begin{array}{l}
P(f,t,m_1,\ldots,m_7)
=f^4-m_1tf^3+(m_2t^2-3-t^8)f^2\\[1mm]
\hskip30mm +(m_7t^7-m_3t^3+2m_1t)f+(t^8-m_6t^6+m_4t^4-m_2t^2+1), 
\end{array}
\end{equation}
and $m_k$ ($k=1,2,\ldots 7$) are the elementary symmetric functions of
$k$-th degree in $b_i~(i=1,2,\ldots,8)$ with
\begin{equation}
b_1b_2\cdots b_8=1. 
\end{equation}
Moreover,
\begin{equation}
\ol{t}=qt,~t=q^{\hf}s. 
\end{equation}
 \item Constraint on parameters \cite{MSY}
\begin{equation}
qb_1b_3b_5b_7=1, \quad b_2b_4b_6b_8=q. \label{qPe8:par}
\end{equation}
\item Hypergeometric solution
\begin{equation}
z=\frac{g-\left(\dfrac{s}{b_1}+\dfrac{b_1}{s}\right)}
     {g-\left(\dfrac{s}{b_8}+\dfrac{b_8}{s}\right)}
=\lambda~\frac{\Phi(q^4a_0;a_1,q^2a_2,\ldots,q^2a_7)}
              {\Phi(a_0;a_1\ldots,a_7)},
\end{equation}
where $\Phi$ is defined in terms of the balanced ${}_{10}W_9$ series by
\begin{equation}
\begin{array}{rl}
&\Phi(a_0;a_1,\ldots,a_7)
 ={}_{10}W_9(a_0;a_1,\ldots,a_7;q^2,q^2)\\[1mm]
 &\hskip2mm
+\dfrac{\left(q^2a_0,\dfrac{a_7}{a_0};q^2\right)_{\infty}}
       {\left(\dfrac{a_0}{a_7},\dfrac{q^2a_7^2}{a_0};q^2\right)_{\infty}}
{\displaystyle \prod_{k=1}^6}\dfrac{\left(a_k,\dfrac{q^2a_7}{a_k};q^2\right)_{\infty}}
                   {\left(\dfrac{q^2a_0}{a_k},\dfrac{a_ka_7}{a_0};q^2\right)_{\infty}}
~{}_{10}W_9\left(\dfrac{a_7^2}{a_0};
                  \dfrac{a_1a_7}{a_0},\ldots,\dfrac{a_6a_7}{a_0},a_7
                  ;q^2,q^2\right).
\end{array} \label{e8:hyper}
\end{equation}
Here, $a_i~(i=0,1,\ldots,7$) and $\lambda$ are given by
\begin{equation}\label{qPe8:param}
\begin{array}{l}
a_0=\dfrac{1}{qb_1b_2b_8^2},\quad 
a_1=\dfrac{q^2}{b_2b_8t^2}, \quad 
a_2=\dfrac{s^2}{b_2b_8}, \\[4mm]
a_i=\dfrac{b_i}{b_8}\ (i=3,5,7),\quad a_i=\dfrac{b_i}{b_1}\ (i=4,6), 
\end{array}
\end{equation}
and
\begin{equation}
\begin{array}{l}
\lambda=
\dfrac{b_1b_4b_6}{b_8s^2}\,
\dfrac{\left(1-\dfrac{b_4b_6}{b_1b_8}\right)
       \left(1-q^2\dfrac{b_4b_6}{b_1b_8}\right)
       (1-b_3b_5t^2)(1-b_3b_7t^2)(1-b_5b_7t^2)
       {\displaystyle \prod_{i=2,4,6}}\left(1-\dfrac{b_i}{b_1}\right)}
      {\left(1-\dfrac{s^2}{b_1b_8}\right)
       \left(1-\dfrac{q^2s^2}{b_1b_8}\right)
       \left(1-\dfrac{b_4}{b_8}\right)
       \left(1-\dfrac{b_6}{b_8}\right)
       \left(1-\dfrac{q}{b_1b_8s^2}\right)
       {\displaystyle \prod_{i=3,5,7}}\left(1-\dfrac{b_4b_6}{b_1b_i}\right)}, 
\end{array}
\end{equation}
respectively.
\end{enumerate}
\subsubsection{Data}
\begin{enumerate}
 \item Riccati equation
\begin{equation}
\left|
\begin{array}{cccc}
1&f&g&fg\\
1&f_1&g_1&f_1g_1\\
1&f_3&g_3&f_3g_3\\
1&f_5&g_5&f_5g_5
\end{array}
\right|=0, \quad 
\left|
\begin{array}{cccc}
1&f&\ol{g}&f\ol{g}\\
1&f_8&\ol{g}_8&f_8\ol{g}_8\\
1&f_6&\ol{g}_6&f_6\ol{g}_6\\
1&f_4&\ol{g}_4&f_4\ol{g}_4
\end{array}
\right|=0, 
\end{equation}
where
 \begin{equation}
f_i=b_it+\frac{1}{b_it}, \quad g_i=\frac{s}{b_i}+\frac{b_i}{s}. \label{e8:fi_gi}
\end{equation}
The Riccati equation for ${\displaystyle z=\frac{g-g_1}{g-g_8}}$ is given by
\begin{equation}
\overline{z}=\frac{Az+B}{Cz+D}, 
\end{equation}
\begin{equation}
\begin{array}{l}
B=-f_{35}g_{13}g_{15}d'_{1468}, \\[1mm]
C= f_{46}\overline{g}_{48}\overline{g}_{68}d_{1358}, \\[1mm]
D=-f_{35}f_{46}f_{18}g_{13}g_{15}\overline{g}_{48}\overline{g}_{68}, \\[1mm]
\Delta=AD-BC=f_{13}f_{35}f_{15}f_{46}f_{48}f_{68}
          g_{13}g_{15}g_{35}g_{18}
          \overline{g}_{46}\overline{g}_{48}\overline{g}_{68}\overline{g}_{18},
\end{array}
\end{equation}
where $f_{ij}=f_i-f_j$ and
\begin{equation}
d_{1358}=
\left|
\begin{array}{cccc}
1&f_1&g_1&f_1g_1\\
1&f_3&g_3&f_3g_3\\
1&f_5&g_5&f_5g_5\\
1&f_8&g_8&f_8g_8
\end{array}
\right|, \quad 
d'_{1468}=
\left|
\begin{array}{cccc}
1&f_1&\overline{g}_1&f_1\overline{g}_1\\
1&f_4&\overline{g}_4&f_4\overline{g}_4\\
1&f_6&\overline{g}_6&f_6\overline{g}_6\\
1&f_8&\overline{g}_8&f_8\overline{g}_8
\end{array}
\right|, 
\end{equation}
respectively.
 \item Three-term and contiguity relations for hypergeometric function \cite{Gupta-Masson}
\begin{enumerate}
 \item Three-term relation
\begin{equation}
{\displaystyle U_1(\overline{\Phi}-\Phi)+U_2\Phi+U_3(\underline{\Phi}-\Phi)=0},  
\end{equation}
where
\begin{equation}
\begin{array}{l}
\medskip
{\displaystyle U_1=
\dfrac{a_1(1-a_2)\left(1-\dfrac{a_0}{a_2}\right)\left(1-\dfrac{q^2a_0}{a_2}\right)}
{(1-\dfrac{q^2a_2}{a_1}){\displaystyle \prod_{j=3}^7}\left(1-\dfrac{q^2a_0}{a_1a_j}\right)},}
\\
{\displaystyle U_2=-(a_1-a_2)\left(1-\frac{q^2a_0}{a_1a_2}\right)\prod_{j=3}^7(1-a_j),}
 \quad
{\displaystyle U_3=U_1|_{a_1\leftrightarrow a_2},}
\end{array}
\end{equation}
\begin{equation}
 \overline{\Phi}=\Phi(a_0;a_1/q^2,q^2a_2,a_3,\ldots,a_7),\quad
 \underline{\Phi}=\Phi(a_0;q^2a_1,a_2/q^2,a_3,\ldots,a_7),
\end{equation}
and $\Phi$ is defined by eq.(\ref{e8:hyper}).
 \item Contiguity relations
\begin{eqnarray}
&& \Phi(a_0;a_1/q^2,q^2a_2,a_3,\ldots,a_7)-
 \Phi(a_0;a_1,a_2,a_3,\ldots,a_7)\nonumber\\
&&=V_1~\Phi(q^4a_0^2;a_1,q^2a_2,\ldots,q^2a_7),\\[2mm]
&&V_2~\Phi(q^4a_0^2;a_1,q^2a_2,q^2a_3,\ldots,q^2a_7)-V_3~\Phi(q^4a_0^2;q^2a_1,a_2,q^2a_3,\ldots,q^2a_7)
\nonumber\\
&&=V_4~\Phi(a_0;a_1,a_2,a_3,\ldots,a_7),
\end{eqnarray}
where
\begin{equation}
\begin{array}{l}
\medskip
V_1= \dfrac{\dfrac{q^2a_0}{a_2}
       \left(1-\dfrac{q^2a_2}{a_1}\right)\left(1-\dfrac{a_1a_2}{q^2a_0}\right)
       (1-q^2a_0)(1-q^4a_0){\displaystyle \prod_{j=3}^7}(1-a_j)}
      {\left(1-\dfrac{q^2a_0}{a_1}\right)\left(1-\dfrac{q^4a_0}{a_1}\right)
       \left(1-\dfrac{a_0}{a_2}\right)\left(1-\dfrac{q^2a_0}{a_2}\right)
       {\displaystyle \prod_{j=3}^7}\left(1-\dfrac{q^2a_0}{a_j}\right)}, \\
\medskip
V_2=\dfrac{ a_1^2(1-a_2){\displaystyle \prod_{j=3}^7}\left(1-\dfrac{q^2a_0}{a_1a_j}\right)}
        {\left(1-\dfrac{q^2a_0}{a_1}\right)\left(1-\dfrac{q^4a_0}{a_1}\right)},\quad
V_3=V_2|_{a_1\leftrightarrow a_2},\\
V_4=\dfrac{ a_1\left(1-\dfrac{a_2}{a_1}\right)
         {\displaystyle \prod_{j=3}^7}\left(1-\dfrac{q^2a_0}{a_j}\right)}
        {(1-q^2a_0)(1-q^4a_0)}.
\end{array}
\end{equation}
\end{enumerate}
 \item Decoupling factors
\begin{equation}
 H=\frac{D}{\Delta}=
-\frac{f_{18}}
{f_{13}f_{15}f_{48}f_{68}g_{35}g_{18}\overline{g}_{46}\overline{g}_{18}},\quad
K=\frac{1}{D}=-\frac{1}{f_{35}f_{46}f_{18}g_{13}g_{15}\overline{g}_{48}\overline{g}_{68}},
\end{equation}
so that
\begin{equation}
 k =\frac{H}{K}=
  \frac{D^2}{\Delta}=\frac{f_{35}f_{46}f_{18}^2g_{13}g_{15}\ol{g}_{48}\ol{g}_{68}}
{f_{13}f_{15}f_{48}f_{68}g_{35}\ol{g}_{46}\ol{g}_{18}}.
\end{equation}
 \item Identification
\begin{equation}
 z=\frac{1}{\kappa}~\frac{F}{G},\quad
F\propto \Phi(q^4a_0;a_1,q^2a_2,\ldots,q^2a_7),\quad
G\propto \Phi(a_0;a_1\ldots,a_7),\quad \ol{\kappa}=k\kappa,
\end{equation}
where $a_i$ ($i=0,\ldots,7$) are given by eq.(\ref{qPe8:param}).
 \item Gauge factors\\
Putting
\begin{equation}
\begin{array}{l}
\smallskip
{\displaystyle F=\theta(q^4a_0;a_1,q^2a_2,\ldots,q^2a_7)\Phi(q^4a_0;a_1,q^2a_2,\ldots,q^2a_7),}
 \\
{\displaystyle G=\theta(a_0;a_1\ldots,a_7)\Phi(a_0;a_1\ldots,a_7),}
\end{array}
\end{equation}
we have:
\begin{equation}
 \frac{\theta(a_0;a_1/q^2,a_2q^2,\ldots,a_7)}{\theta(a_0;a_1\ldots,a_7)}=1,\quad
 \frac{1}{\kappa}\frac{\theta(q^4a_0;a_1,q^2a_2,\ldots,q^2a_7)}{\theta(a_0;a_1\ldots,a_7)}
=\lambda.
\end{equation}
\end{enumerate}
\subsection{Case of $E_6^{(1)}$}
\subsubsection{Equation and Solution}
\begin{enumerate}
 \item $q$-Painlev\'e equation\cite{RGTT,GROe6geo,TRGOe6sol}
\begin{equation}
\begin{array}{l}\medskip
{\displaystyle (f\og - 1)(fg-1) = t\ot~\frac{(f-b_1t)(f-b_2t)(f-b_3t)(f-b_4 t)}
{(f-b_5t)\left(f-\dfrac{t}{b_5}\right)}},\\
\medskip
{\displaystyle (fg - 1)(\df g-1) = 
t^2~\frac{\left(g-\dfrac{1}{b_1}\right)\left(g-\dfrac{1}{b_2}\right)\left(g-\dfrac{1}{b_3}\right)
\left(g-\dfrac{1}{b_4}\right)}
{\left(g-b_6t\right)\left(g-\dfrac{t}{b_6}\right)}},\\
\ot=qt,\quad b_1b_2b_3b_4=1.
\end{array}
\label{qPe6}
\end{equation} 
 \item Constraint on parameters \cite{RGTT,TRGOe6sol}
\begin{equation}
b_1b_2=b_5b_6 .
\end{equation}
 \item Hypergeometric solution
\begin{equation}
z=\frac{g-\dfrac{1}{b_1}}{g-tb_6}=\frac{1-\dfrac{b_3}{b_1}}{1-\dfrac{b_1b_2b_3t}{b_5}}~
\frac{\Phi(qa,b,c,d,e)}{\Phi(a,b,qc,d,e)},\label{qPe6:hyper}
\end{equation}
where $\Phi$ is the balanced ${}_3\varphi_2$ series defined by
\begin{equation}
\Phi(a,b,c,d,e)={}_3\varphi_2\left(\begin{array}{c}a,b,c \\d,e\end{array};q,\frac{de}{abc}\right),
\end{equation}
with
\begin{equation}
  a=\frac{b_3b_5}{t},\quad b=\frac{b_3}{b_2},\quad c=b_1^2b_2b_3,\quad d=q\frac{b_3b_5^2}{b_2},\quad
e=qb_1b_2b_3^2.\label{qPe6:param}
\end{equation}
\end{enumerate}
\subsubsection{Data}
\begin{enumerate}
 \item Riccati equation \cite{RGTT,TRGOe6sol}
\begin{equation}
\og=\frac{1+\dfrac{b_5\ot}{b_1b_2} (f-b_1-b_3)}{f-\ot b_5},\quad
f=\frac{1+b_6 t(b_3b_4g-b_3-b_4)}{g-tb_6}.
\label{qPe6:ric1}
\end{equation}
The Riccati equation for
\begin{equation}
z=\frac{g-1/b_1}{g-tb_6},
\end{equation}
is given by,
\begin{displaymath}
 \overline{z}=\frac{Az+B}{Cz+D},
\end{displaymath}
\begin{equation}
\begin{array}{l}
\smallskip
A=b_1b_5(b_3b_5 - t)(b_5 - b_1 b_2 b_3 t) (-b_2 + b_5 q t),\\
\smallskip
B=-b_5^2 t (b_1 - b_3)(-1 + b_1^2 b_2 b_3) (-b_2 + b_5 q t),\\
\smallskip
C=qtb_1 (b_1 b_2 - b_5) (b_1 b_2 + b_5) (b_3 b_5 - t) (-b_5 + b_1 b_2 b_3 t),\\
D=-b_5\Bigl[b_1b_2b_3b_5^2 +(- b_1^3b_2^2b_3b_5 -  b_1^3b_2^2b_3b_5q - b_2b_3b_5^3q)t\\
\hskip10pt 
+ (b_1^3b_2^2 - b_1^2b_2^2b_3+ b_1^4b_2^3b_3^2 - b_1b_5^2 +  b_3b_5^2 + b_1^3b_2b_3b_5^2 + 
        b_1^2b_2^2b_3b_5^2 - b_1^2b_2b_3^2b_5^2+ b_1^2b_2^2b_3b_5^2q)qt^2\\
\hskip10pt -  b_1^4b_2^3b_3b_5q^2t^3\Bigr].
\end{array}
\end{equation}
 \item Three-term and contiguity relations for hypergeometric function \cite{Gupta-Ismail-Masson}
\begin{enumerate}
 \item Three-term relation
\begin{equation}
V_1\left(\overline{\Phi}-\Phi\right)
+V_2\Phi + 
V_3\left(\Phi-\underline{\Phi}\right)=0,
\end{equation}
where
\begin{equation}
\begin{array}{l}
\smallskip
{\displaystyle V_1=\left(1-\frac{q}{z}\right)\left(1-a\right), 
\quad V_2=(1-b)(1-c),\quad
V_3=\frac{a}{z}\left(1-\frac{d}{a}\right)\left(1-\frac{e}{a}\right)},\\
{\displaystyle  \Phi=\Phi(a,b,c,d,e)={}_3\varphi_2\left(\begin{array}{c}a,b,c\\d,e
			 \end{array};q;z\right),\quad z=\frac{de}{abc},
\quad \overline{\Phi}=\left.\Phi\right|_{a\to qa},
\quad \underline{\Phi}=\left.\Phi\right|_{a\to a/q}.}
\end{array}
\end{equation}
 \item Contiguity relations
 \begin{eqnarray}
&&(a-c)\Phi(a,b,c,d,e) +
 (1-a)\Phi(qa,b,c,d,e)-(1-c)\Phi(a,b,qc,d,e)=0,\\[2mm]
&& (a-c)(de-abc)\Phi(a,b,c,d,e) + bc(d-a)(e-a)\Phi(a/q,b,c,d,e)\nonumber\\
&&-ab(d-c)(e-c)\Phi(a,b,c/q,d,e)=0.
 \end{eqnarray}
\end{enumerate}
 \item Decoupling factors
\begin{equation}
 \begin{array}{l}
\medskip
{\displaystyle H=\frac{1}{b_1b_3b_5(-b_5 + b_1^2 b_2  t)(b_5 -
b_1b_2b_3t)(b_2 -   qb_5t)},}\\
\medskip
{\displaystyle K=\frac{1}{b_1b_3b_5(-b_5 + b_1^2 b_2  t)(b_5 -
qb_1b_2b_3t)(b_2 -  qb_5t)},}\\
{\displaystyle k=\frac{H}{K}=\frac{b_5-qb_1b_2b_3t}{b_5-b_1b_2b_3t},
 \quad \kappa=1-\frac{b_1b_2b_3t}{b_5}}.
 \end{array}
\end{equation}
 \item Identification
\begin{equation}
 z=\frac{1}{\kappa}~\frac{F}{G},\quad
F\propto \Phi(qa,b,c,d,e),\quad G\propto \Phi(a,b,qc,d,e),
\end{equation}
where $a,\ldots,e$ are given by eq.(\ref{qPe6:param}).
 \item Gauge factors\\
Putting
\begin{equation}
 F=\theta(qa,b,c,d,e)\Phi(qa,b,c,d,e),\quad
 G=\theta(a,b,qc,d,e)\Phi(a,b,qc,d,e),
\end{equation}
we have:
\begin{equation}
\frac{\theta(a,b,c,d,e)}{\theta(a,b,qc,d,e)}=\frac{\theta(qa,b,c,d,e)}{\theta(a,b,qc,d,e)}
=1-\dfrac{b_3}{b_1},\quad\frac{\theta(a/q,b,qc,d,e)}{\theta(a,b,qc,d,e)}=1.
\end{equation}
\end{enumerate}
\subsection{Case of $A_4^{(1)}$}
\subsubsection{Equation and Solution}
\begin{enumerate}
 \item $q$-Painlev\'e equation \cite{RGTT}
\begin{equation}
\begin{array}{l}
\medskip
{\displaystyle \ol{g}g=\frac{\left(f+\dfrac{a_1}{t}\right)\left(f+\dfrac{1}{a_1t}\right)}{1+a_3f}},\\
\medskip
{\displaystyle f\ul{f}=\frac{\left(g+\dfrac{a_2}{s}\right)\left(g+\dfrac{1}{a_2s}\right)}{1+g/a_3},}\\
\ot=qt,\quad t=q^{\frac{1}{2}}s.
\end{array}
\label{qPa4}
\end{equation} 
 \item Constraint on parameters \cite{RGTT}
\begin{equation}
a_1a_2a_3^2=q^{-\frac{1}{2}}.
\end{equation}
 \item Hypergeometric solution
\begin{equation}
g=-\frac{1}{a_1a_3^2t}~\frac{\Phi(\alpha_1,\alpha_2,z)}{\Phi(\alpha_1,q\alpha_2,z)},\quad
f=\frac{1}{a_3}\left(1-\frac{1}{a_2^2}\right)~
\frac{\Phi(q\alpha_1,q\alpha_2,z)}{\Phi(\alpha_1,q\alpha_2,z)},
\label{qPa4:hyper}
\end{equation}
where $\Phi$ is the  ${}_2\varphi_1$ series defined by
\begin{equation}
 \Phi(\alpha_1,\alpha_2,z)={}_2\phi_1\left(\begin{array}{c}\alpha_1,\alpha_2\\0\end{array};q,z\right),
\end{equation}
with
\begin{equation}
\alpha_1=\frac{1}{a_2^2},\quad \alpha_2=a_1^2,\quad z=\frac{t}{a_1a_3}.\label{qPa4:param}
\end{equation}
Note that the solution is also expressible in terms of ${}_1\phi_1$
       series by using the formula \cite{Koekoek-Swarttouw},
\begin{equation}
 {}_2\phi_1\left(\begin{array}{c}a,b\\0\end{array};q,z\right)=\frac{(bz;q)_\infty}{(z;q)_\infty}
~{}_1\phi_1\left(\begin{array}{c}b\\bz\end{array};q,az\right).
\end{equation}
\end{enumerate}
\subsubsection{Data}
\begin{enumerate}
 \item Riccati equation \cite{RGTT}
\begin{equation}
\ol{g}=\frac{a_3^2~g+\dfrac{1-a_1^2}{a_1t}}{-a_3g+\left(\dfrac{1}{a_3^2}-\dfrac{1}{a_1a_3t}\right)},\quad 
g=-\frac{f+\dfrac{1}{a_1t}}{a_3^2}.
\end{equation}
 \item Three-term and contiguity relations for hypergeometric function
\begin{enumerate}
 \item Three-term relation
\begin{equation}
 \frac{\alpha_1\alpha_2}{q}z~\left(\ol{\Phi}-\Phi\right)
+\frac{z}{q}(1-\alpha_1)(1-\alpha_2)~\Phi
+\left(\frac{z}{q}-1\right)~\left(\ul{\Phi}-\Phi\right)=0,
\end{equation}
where
\begin{equation}
 \Phi(\alpha_1,\alpha_2,z)={}_2\phi_1\left(\begin{array}{c}\alpha_1,\alpha_2\\0\end{array};q,z\right),\quad
 \ol{\Phi}=\left.\Phi\right|_{z\to qz},\quad \ul{\Phi}=\left.\Phi\right|_{z\to z/q}.
\end{equation}
 \item Contiguity relations
\begin{eqnarray}
&&\Phi(\alpha_1,\alpha_2,z)-\alpha_2\Phi(\alpha_1,\alpha_2,qz)=
(1-\alpha_2)~\Phi(\alpha_1,q\alpha_2,z), \\
&&\Phi(\alpha_1,\alpha_2/q,z)-\Phi(\alpha_1,\alpha_2,z)=
-\frac{\alpha_2z}{q}(1-\alpha_1)~\Phi(q\alpha_1,\alpha_2,z),\\
&& \alpha_1\alpha_2z~\Phi(\alpha_1,q\alpha_2,qz)=\Phi-(1-\alpha_2z)~\Phi(\alpha_1,q\alpha_2,z).
\end{eqnarray}
\end{enumerate}
\item Decoupling factors
\begin{equation}
H=\frac{1}{a_1^2a_3^2},\quad K=\frac{1}{qa_1^2a_3^2},\quad
 k=\frac{H}{K}=q,\quad \kappa=t.
\end{equation}
\item Identification
\begin{equation}
 f=\frac{1}{\kappa}~\frac{F}{G},\quad F\propto
  \Phi(\alpha_1,\alpha_2,z),\quad G\propto \Phi(\alpha_1,q\alpha_2,z),
\end{equation}
with parameters given in eq.(\ref{qPa4:param}).
\item Gauge factors

Putting
\begin{equation}
 F=\theta(\alpha_1,\alpha_2,z)\Phi(\alpha_1,\alpha_2,z),\quad
 G=\theta(\alpha_1,q\alpha_2,z)\Phi(\alpha_1,q\alpha_2,z),
\end{equation}
we have:
\begin{equation}
\frac{\theta(\alpha_1,\alpha_2,qz)}{\theta(\alpha_1,\alpha_2,z)}=1,\quad
\frac{\theta(\alpha_1,q\alpha_2,z)}{\theta(\alpha_1,\alpha_2,z)}=-a_1a_3^2.
\end{equation}
\end{enumerate}
\subsection{Case of $(A_2+A_1)^{(1)}$}
\subsubsection{Equation and Solution}
\begin{enumerate}
 \item $q$-Painlev\'e equation \cite{Sakai,RGTT,KNY:qp4,KK:qp3,K:qp3}
\begin{equation}
 \ol{g}gf=b_0~\frac{1+a_0tf}{a_0t+f},\quad
gf\ul{f}=b_0~\frac{\dfrac{a_1}{t}+g}{1+\dfrac{a_1}{t}g},\quad \ot =qt.\label{qPa2a1}
\end{equation}
Eq. (\ref{qPa2a1}) admits two different specializations for
hypergeometric solutions: (a) specialization of $b_i$ (parameter of
       $A_1$), (b) specialization of $a_i$ (parameter of $A_2$). See also
       \cite{KNY:qp4,KK:qp3} for details.
\item Constraint on parameters
\begin{enumerate}
 \item 
\begin{equation}
 b_0=q.
\end{equation}
 \item 
\begin{equation}
a_0a_1=q.
\end{equation}
\end{enumerate}
 \item Hypergeometric solution
\begin{enumerate}
 \item 
\begin{equation}
  g=-\frac{a_1}{t}\left(1-\frac{q^2}{a_0^2a_1^2}\right)~\frac{\Phi(b,z)}{\Phi(q^2b,z)},\quad
  f=\frac{q^2t}{a_0a_1^2}~\frac{1}{1-\dfrac{q^2}{a_0^2a_1^2}}~\frac{\Phi(q^2b,q^2z)}{\Phi(b,z)},
\label{qPa2a1:hyper1}
\end{equation}
where 
\begin{equation}
 \Phi=\Phi(b,z)={}_1\varphi_1\left(\begin{array}{l}0 \\b\end{array};q^2,z\right),
\end{equation}
with
\begin{equation}
b=q^2/a_0^2a_1^2,\quad z=q^2t^2/a_1^2.\label{qPa2a1:param1}
\end{equation}
 \item 
\begin{equation}
  g=\frac{b_0}{a_0t}~\frac{\Psi(a,z)}{\Psi(a,q^2z)},\quad
f=-a_0t~\frac{\Psi(a,q^2z)}{\Psi(a,z)},\label{qPa2a1:hyper2}
\end{equation}
where 
\begin{equation}
\Psi=\Psi(a,z)= {}_1\varphi_1\left(\begin{array}{l}a \\0\end{array};q^2,z\right),
\end{equation}
with 
\begin{equation}
a=a_0^2t^2,\quad z=q/b_0.\label{qPa2a1:param2}
\end{equation}
\end{enumerate}
\end{enumerate}
\subsubsection{Data}
\begin{enumerate}
 \item Riccati equation
\begin{enumerate}
 \item 
\begin{equation}
\overline{f}=\frac{\left(a_0^2a_1^2/q-a_0^2qt^2\right)f-a_0qt}{a_0tf+1},\quad
g=-a_0a_1~\frac{1}{a_0t+f}.
\end{equation}
 \item 
\begin{equation}
 \overline{g}=-\frac{g-a_0b_0t}{a_0t g-b_0},\quad fg=-b_0.
\end{equation}
\end{enumerate}
 \item Three-term and contiguity relations for hypergeometric
    function
\begin{enumerate}
 \item 
\begin{enumerate}
 \item Three-term relation
\begin{equation}
 \frac{b}{z}\left(\Phi(b,q^2z)-\Phi(b,z)\right)+\Phi(b,z)+\frac{q^2}{z}\left(\Phi(b,z/q^2)-\Phi\right)=0,
\end{equation}
where
\begin{equation}
 \Phi(b,z)= {}_1\varphi_1\left(\begin{array}{l}0 \\b\end{array};q^2,z\right).
\end{equation}
 \item Contiguity relations
 \begin{eqnarray}
&& \Phi(b,z)-\frac{b}{q^2}\Phi(b,q^2z)=\left(1-\frac{b}{q^2}\right)\Phi(b/q^2,z),\\
&& \Phi(b,z)-\Phi(b,z/q^2)=\frac{z/q^2}{1-b}~\Phi(q^2b,z).
\end{eqnarray}
\end{enumerate}
 \item 
\begin{enumerate}
 \item Three-term relation
\begin{equation}
(1-a)\frac{z}{q^2}\left(\Psi(q^2a,z)-\Psi(a,z)\right)-\frac{az}{q^2}\Psi(a,z) 
+ \left(\Psi(a/q^2,z)-\Psi(a,z)\right)=0,
\end{equation}
where 
\begin{equation}
 \Psi(a,z)={}_1\varphi_1\left(\begin{array}{l}a \\0\end{array};q^2,z\right).
\end{equation}
 \item Contiguity relations
\begin{eqnarray}
&&\Psi(a,z)-a\Psi(a,q^2z)=(1-a)\Psi(q^2a,z),\\
&&\Psi(a,z)-\Psi(q,z/q^2)=(1-a)\frac{z}{q^2}~\Psi(q^2a,z).
\end{eqnarray}
\end{enumerate}
\end{enumerate}
\item Decoupling factors
\begin{enumerate}
 \item 
\begin{equation}
 H=\frac{1}{q},\quad K=1,\quad k=\frac{H}{K}=\frac{1}{q},\quad \kappa=\frac{a_1}{qt}.
\end{equation}
 \item 
\begin{equation}
 H=-\frac{1}{1-a_0^2t^2},\quad K=\frac{1}{q}~\frac{1}{1-a_0^2t^2},\quad
k=\frac{H}{K}=q,\quad \kappa=a_0t.
\end{equation}
\end{enumerate}
\item Identification
\begin{enumerate}
 \item 
\begin{equation}
 f=\frac{1}{\kappa}~\frac{F}{G},\quad F\propto \Phi(q^2b,q^2z),\quad G\propto \Phi(b,z),
\end{equation}
with parameters given in eq.(\ref{qPa2a1:param1}).
 \item 
\begin{equation}
 g=\frac{1}{\kappa}~\frac{F}{G},\quad F\propto \Psi(a,z),\quad G\propto \Psi(a,q^2z),
\end{equation}
with parameters given in eq.(\ref{qPa2a1:param2}).
\end{enumerate}
\item Gauge factors
\begin{enumerate}
 \item Putting
\begin{equation}
 F=\theta(q^2b,q^2z)\Phi(q^2b,q^2z),\quad
 G=\theta(b,z)\Phi(b,z),
\end{equation}
we have:
\begin{equation}
 \frac{\theta(b,q^2z)}{\theta(b,z)}=1,\quad
 \frac{\theta(q^2b,z)}{\theta(b,z)}=\frac{\theta(q^2b,q^2z)}{\theta(b,z)}
=\frac{q}{a_0a_1\left(1-\dfrac{q^2}{a_0^2a_1^2}\right)}.
\end{equation}
 \item Putting
\begin{equation}
 F=\theta(a,z)\Psi(a,z),\quad G=\theta(a,q^2z)\Psi(a,q^2z),
\end{equation}
we have:
\begin{equation}
 \frac{\theta(q^2a,z)}{\theta(a,z)}=1,\quad
\frac{\theta(a,q^2z)}{\theta(a,z)}=\frac{\theta(q^2a,q^2z)}{\theta(a,z)}
=\frac{1}{b_0}.
\end{equation}
\end{enumerate}
\end{enumerate}

\subsection{Case of $(A_1+A_1')^{(1)}$}
\subsubsection{Equation and Solution}
\begin{enumerate}
 \item $q$-Painlev\'e equation \cite{Sakai,RGTT,RG:coales}
\begin{equation}
(\overline{f}f-1)(f\underline{f}-1)=\frac{at^2f}{f+t},\quad \overline{t}=qt.\label{qPa1a1}
\end{equation}
 \item Constraint on parameters
\begin{equation}
 a=q.
\end{equation}
 \item Hypergeometric solution
\begin{equation}
f=\frac{\Phi(qt)}{\Phi(t)},\quad  \Phi={}_1\varphi_1\left(\begin{array}{c}  0\\-q \end{array};q,-qt\right).
\end{equation}
\end{enumerate}
\subsubsection{Data}
\begin{enumerate}
 \item Riccati equation
\begin{equation}
  \overline{f}=\frac{1}{f}-qt.
\end{equation}
 \item Three-term relation
\begin{equation}
 \Phi(qt)+t\Phi(t)=\Phi(t/q).
\end{equation}
 \item Identification
\begin{equation}
f= \frac{F}{G},\quad  F=\Phi(qt),\quad G=\Phi(t).
\end{equation}
\end{enumerate}
We note that there is no need to introduce decoupling and gauge factors.\\

\noindent\textbf{Acknowledgment} One of the author(K.K.) would like to
thank Professor Nalini Joshi for hospitality during his stay in
University of Sydney.


\end{document}